\documentclass{PoS}
\usepackage{amsmath}
\usepackage{axodraw}

\def\beq{\begin{equation}}
\def\eeq{\end{equation}}
\def\bea{\begin{eqnarray}}
\def\eea{\end{eqnarray}}

\def\nn{\nonumber}
\def\Eq#1{Eq.~(\ref{#1})}

\def\pb{p\hspace{-.42em}/\hspace{-.07em}}
\def\qb{q\hspace{-.42em}/\hspace{-.07em}}
\def\td#1{\tilde{\delta}\left(#1\right)}

\newcommand{\la}{\langle}
\newcommand{\ra}{\rangle}

\def\sp{{\mathbf{Sp}}}
\def\ket#1{|{#1}\ra}
\def\bra#1{\la{#1}|}
\def\qb{\mathbf{q}}
\def\pb{\mathbf{p}}
\title{The loop-tree duality at work}

\ShortTitle{The loop-tree duality at work}

\author{Sebastian Buchta, Grigorios Chachamis, Ioannis Malamos \\
Instituto de F\'{\i}sica Corpuscular, UVEG - Consejo Superior de 
Investigaciones Cient\'{\i}ficas, \\ 
Parc Cient\'{\i}fic, E-46980 Paterna, Valencia, Spain \\
E-mail: \email{sebastian.buchta@ific.uv.es, grigorios.chachamis@ific.uv.es,
ioannis.malamos@ific.uv.es}}

\author{Isabella Bierenbaum \\
II. Institut f\"ur Theoretische Physik, Universit\"at Hamburg,
Luruper Chaussee 149, 22761 Hamburg, Germany \\
E-mail: \email{isabella.bierenbaum@desy.de}}

\author{Petros Draggiotis \\
Institute of Nuclear and Particle Physics, NCSR Demokritos, 
Agia Paraskevi, 15310, Greece \\
E-mail: \email{petros.draggiotis@gmail.com}}

\author{\speaker{Germ\'an Rodrigo} \\
Instituto de F\'{\i}sica Corpuscular, UVEG - Consejo Superior de 
Investigaciones Cient\'{\i}ficas, \\ 
Parc Cient\'{\i}fic, E-46980 Paterna, Valencia, Spain \\
E-mail: \email{german.rodrigo@csic.es}}

\abstract{
We review the recent developments of the loop-tree
duality method, focussing our discussion on analysing 
the singular behaviour of the loop integrand of the dual 
representation of one-loop integrals and scattering amplitudes.
We show that within the loop-tree duality method there is a partial 
cancellation of singularities at the integrand level among the different
components of the corresponding dual representation.
The remaining threshold and infrared singularities are restricted 
to a finite region of the loop momentum space, which is of the 
size of the external momenta and can be mapped to
the phase-space of real corrections to cancel 
the soft and collinear divergences.  
}

\FullConference{Loops and Legs in Quantum Field Theory\\
                 27 April 2014 - 02 May 2014\\
                 Weimar, Germany}

\begin{document}

\section{Introduction}

The recent discovery of the Higgs boson at the LHC represents
a great success of the Standard Model (SM) of elementary particles. 
At the same time, the absence so far of a clear signal 
of physics beyond the SM during the first run of the LHC 
leaves a certain degree of dissatisfaction.
Because of that, the high quality of data that 
the LHC will provide in the next run increases even more 
the relevance of high-precision theoretical 
predictions for the analysis of known phenomena and for finding 
innovative strategies to achieve new discoveries. 

The loop-tree duality method~\cite{Catani:2008xa,Bierenbaum:2010cy,Bierenbaum:2012th,Buchta:2014dfa}
establishes that generic loop quantities 
(loop integrals and scattering amplitudes) 
in any relativistic, local and unitary field theory can be written 
as a sum of tree-level objects obtained after making all possible cuts to 
the internal lines of the corresponding Feynman diagrams, 
with one single cut per loop
and integrated over a measure that closely 
resembles the phase-space of the corresponding real corrections. 
This duality relation is realized by a modification of 
the customary +i0 prescription of the Feynman propagators.
At one-loop, the new prescription compensates for the absence of multiple-cut 
contributions that appear in the Feynman Tree Theorem~\cite{Feynman:1963ax}.
The modified phase-space raises the intriguing possibility
that virtual and real corrections can be brought together under a 
common integral and treated with Monte Carlo techniques at the same time.
In this talk, we review the actual state of development of the loop-tree
duality method and focus our discussion on analysing 
the singular behaviour of the loop integrand of the dual 
representation of one-loop integrals and scattering amplitudes, 
as a necessary step towards a numerical implementation for the 
calculation of physical cross-sections.

\section{The loop-tree duality relation at one-loop}
\label{sec:one-loop}

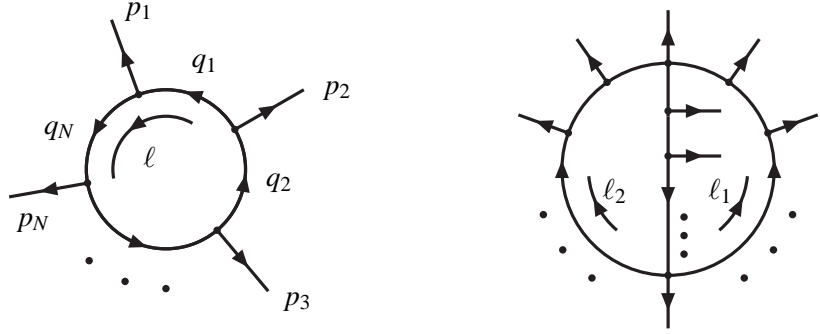
\begin{figure}[t]
\begin{center}
\vspace*{8mm}
\begin{picture}(300,110)(0,-10)
\SetWidth{1.2}
\BCirc(50,50){30}
\ArrowArc(50,50)(30,110,190)
\ArrowArc(50,50)(30,190,-50)
\ArrowArc(50,50)(30,-50,30)
\ArrowArc(50,50)(30,30,110)
\ArrowArc(50,50)(20,60,190)
\ArrowLine(39.74,78.19)(29.48,106.38)
\ArrowLine(75.98,65)(101.96,80)
\ArrowLine(69.28,27.01)(88.56,4.03)
\ArrowLine(20.45,44.79)(-9.09,39.58)
\Vertex(21.07,15.53){1.4}
\Vertex(34.60,7.71){1.4}
\Vertex(50,5){1.4}
\Vertex(39.74,78.19){1.4}
\Vertex(75.98,65){1.4}
\Vertex(69.28,27.01){1.4}
\Vertex(20.45,44.79){1.4}
\Text(44,55)[]{$\ell$}
\Text(40,110)[]{$p_1$}
\Text(65,90)[]{$q_1$}
\Text(115,80)[]{$p_2$}
\Text(93,45)[]{$q_2$}
\Text(10,65)[]{$q_N$}
\Text(0,30)[]{$p_N$}
\Text(100,0)[]{$p_3$}
%%
%%============================================
\SetOffset(190,0)
\ArrowArc(50,50)(40,270,90)
\ArrowArcn(50,50)(40,270,90)
\Line(50,90)(50,70)
\ArrowLine(50,70)(50,10)
\ArrowLine(50,10)(50,-10)
\ArrowLine(50,90)(50,110)
% alpha_1
\ArrowLine(72.9,82.8)(84.4,99.1)
\ArrowLine(87.6,63.7)(106.4,70.52)
\Vertex(72.9,82.8){1.4}
\Vertex(87.6,63.7){1.4}
\Vertex(96.9,32.9){1.4}
\Vertex(89.7,19.6){1.4}
\Vertex(78.7,9){1.4}
% alpha_2
\ArrowLine(27,82.8)(15.6,99.1)
\ArrowLine(12.4,63.7)(-6.4,70.52)
\Vertex(27,82.8){1.4}
\Vertex(12.4,63.7){1.4}
\Vertex(3,32.9){1.4}
\Vertex(10.3,19.6){1.4}
\Vertex(21.3,9){1.4}
% alpha_3
\ArrowLine(50,72)(70,72)
\ArrowLine(50,55)(70,55)
\Vertex(50,72){1.4}
\Vertex(50,55){1.4}
\Vertex(50,90){1.4}
\Vertex(50,10){1.4}
\Vertex(56,32){1.4}
\Vertex(56,25){1.4}
\Vertex(56,18){1.4}
\ArrowArc(50,50)(30,310,-5)
\ArrowArcn(50,50)(30,230,185)
\Text(70,42)[]{$\ell_1$}
\Text(30,42)[]{$\ell_2$}
\end{picture}
\end{center}
\caption{\label{f1loop} 
{\em Momentum configuration of the one-loop (left)
and two-loop (right) $N$-point scalar integrals.}}
\end{figure}

The loop-tree duality relation is obtained by directly 
applying the Cauchy residue theorem to a general one-loop $N$-leg 
scalar integral
\beq
\label{Ln}
L^{(1)}(p_1, \dots, p_N) =
\int_{\ell} \, \prod_{i \in \alpha_1} \, G_F(q_i)~, \quad
\int_{\ell} \bullet = - i \int \frac{d^d \ell}{(2\pi)^{d}} \bullet~,
\eeq
where 
\beq
G_F(q_i)=\frac{1}{q_i^2-m_i^2+i0}
\label{eq:feynman}
\eeq
are Feynman propagators that depend on the 
loop momentum $\ell$, which flows anti-clockwise, 
and the four-momenta of the external legs $p_{i}$, 
$i \in \alpha_1 = \{1,2,\ldots N\}$, which are taken as outgoing and 
are ordered clockwise
(this kinematical configuration is shown in Fig.~\ref{f1loop}(left)). 
We use dimensional regularization with $d$  
the number of space-time dimensions. 
The momenta of the internal lines $q_{i,\mu} = (q_{i,0},\mathbf{q}_i)$, 
where $q_{i,0}$ is the energy (time component) and $\qb_{i}$ are 
the spacial components, are defined as $q_{i} = \ell + k_i$ with 
$k_{i} = p_{1} + \ldots + p_{i}$, and $k_{N} = 0$ by momentum conservation. 
We also define $k_{ji} = q_j - q_i$.

The dual representation of the scalar one-loop integral in \Eq{Ln}
is thus the sum of $N$ dual integrals~\cite{Catani:2008xa,Bierenbaum:2010cy}:
\bea
\label{oneloopduality}
L^{(1)}(p_1, \dots, p_N) 
&=& - \sum_{i\in \alpha_1} \, \int_{\ell} \; \td{q_i} \,
\prod_{\substack{j \in \alpha_1, j\neq i}} \,G_D(q_i;q_j)~,
\eea 
where
\beq
G_D(q_i;q_j) = \frac{1}{q_j^2 -m_j^2 - i0 \, \eta \, k_{ji}}
\eeq
are the so-called dual propagators, as defined in Ref.~\cite{Catani:2008xa},
with $\eta$ a {\em future-like} vector, $\eta^2 \ge 0$, 
with positive definite energy $\eta_0 > 0$.
The delta function 
$\td{q_i} \equiv 2 \pi \, i \, \theta(q_{i,0}) \, \delta(q_i^2-m_i^2)$
sets the internal lines on-shell by selecting the pole of the propagators
with positive energy $q_{i,0}$ and negative imaginary part. 
The presence of the vector
$\eta$ is a consequence of using the residue theorem and the fact that the
residues at each of the poles are not Lorentz-invariant quantities.  The
Lorentz-invariance of the loop integral is recovered after summing over all
the residues. 

%%============================================
%%============================================

\section{Loop-tree duality relation at two-loops and beyond}
\label{sec:two-loops}

The extension of the Duality theorem to two-loops and beyond 
has been discussed in detail in Ref.~\cite{Bierenbaum:2010cy}. 
It is convenient to define the following functions combining different
Feynman and dual propagators:
\beq
\label{eq:multi}
G_{F} ( \alpha_k) = \prod_{i \in \alpha_k} G_{F}( q_i)~, \qquad
G_D( \alpha_k) = \sum_{i \in \alpha_k} \, \td{ q_{i}} \, 
\prod_{\substack{j \in \alpha_k \\ j \neq i }} \, G_D( q_i; q_j)~,
\eeq
where $\alpha_k$ is used to denote any set of internal momenta
that depend on the same loop momentum or the sum of several 
independent loop momenta. At two loops we need three  
\emph{loop lines} $\alpha_k$ to label all the internal momenta: 
$\alpha_1$, $\alpha_2$ and $\alpha_3$ for those momenta that 
depend on $\ell_1$, $\ell_2$ and $\ell_1+\ell_2$, respectively 
(see Fig.~\ref{f1loop}(right)). 
By definition $G_D(\alpha_k)=\td{q_i}$, when $\alpha_k = \{i\}$ 
consists of a single four-momentum. We also define:
\beq
\label{eq:multiminus}
G_D(-\alpha_k) = \sum_{i \in \alpha_k} \, \td{-q_{i}} \, 
\prod_{\substack{j \in \alpha_k \\ j \neq i }} \, G_D(-q_i;-q_j)~, 
\eeq
where the sign in front of $\alpha_k$ indicates that we have reversed the
momentum flow of all the internal lines in $\alpha_k$. 

The key ingredient necessary to extend the loop-tree duality theorem to 
higher orders is the following relationship relating 
the dual and Feynman functions of two subsets: 
\beq
\label{eq:twoGD}
G_D(\alpha_1 \cup \alpha_2) 
= G_D(\alpha_1) \, G_D(\alpha_2)
+ G_D(\alpha_1) \, G_F(\alpha_2)
+ G_F(\alpha_1) \, G_D(\alpha_2)~,
\eeq
which can be generalized as well to the union of an arbitrary 
number of loop lines~\cite{Bierenbaum:2010cy}. The application of the
loop-tree duality theorem at higher orders proceeds in a recursive way.  
For the two-loop case, one starts by selecting one of the loops
\bea
L^{(2)}(p_1, p_2, \dots, p_N) &=& \int_{\ell_1} \, \int_{\ell_2} \,  
G_F(\alpha_1 \cup \alpha_2 \cup \alpha_3)  \nn \\
&=& - \int_{\ell_1} \, \int_{\ell_2} \,  
G_F(\alpha_2) \, G_D(\alpha_1 \cup \alpha_3)~. 
\eea
As the loop-tree duality theorem applies to Feynman propagators only, 
we use~\Eq{eq:twoGD} to re-express the dual propagators entering 
the second loop as Feynman propagators. The application of the 
loop-tree duality theorem to the second loop with momentum $\ell_2$
also requires to reverse the momentum flow in some of the loop lines. 
The final dual representation of a two-loop scalar integral reads:
\bea
\label{AdvDual}
L^{(2)}(p_1, p_2, \dots, p_N) &=&   \int_{\ell_1} \int_{\ell_2} \, \{
- G_D(\alpha_1) \, G_F(\alpha_2) \, G_D(\alpha_3) \\
&+& G_D(\alpha_1) \, G_D(\alpha_2\cup \alpha_3)
+ G_D(\alpha_3) \, G_D(-\alpha_1\cup \alpha_2) \}~, \nn 
\eea 
which is given by double cut contributions opening 
the loop diagram to a tree-level object.

\section{The loop-tree duality relation for multiple poles}
\label{sec:multiple}

The appearance of identical propagators or powers of propagators 
can be avoided at one-loop by a convenient 
choice of the gauge~\cite{Catani:2008xa}, but not at higher orders.  
Identical propagators possess higher than single poles and the loop-tree
duality theorem discussed so far, which is based on assuming single poles, 
must be extended to accommodate for this new feature. 
Two different strategies have been proposed in Ref.~\cite{Bierenbaum:2012th}
to deal with this problem. The first one consists of extending the 
loop-tree duality theorem by using the Cauchy residue theorem for 
higher order poles. 
The second one consists of using Integration by Parts 
(IBP)~\cite{Chetyrkin,Smirnov} to reduce integrals with multiple poles 
to integrals with single poles where the original loop-tree duality theorem
can be applied directly. It is important to stress
that in that case it is not necessary to perform a full reduction to a 
particular integral basis. Explicit examples at two- and three-loops 
have been presented in Ref.~\cite{Bierenbaum:2012th}.

\section{Cancellation of singularities among dual integrands}
\label{sec:cancel}

%%%%%%%%%%%%%%%
\begin{figure}[th]
\begin{center}
\includegraphics[width=6cm]{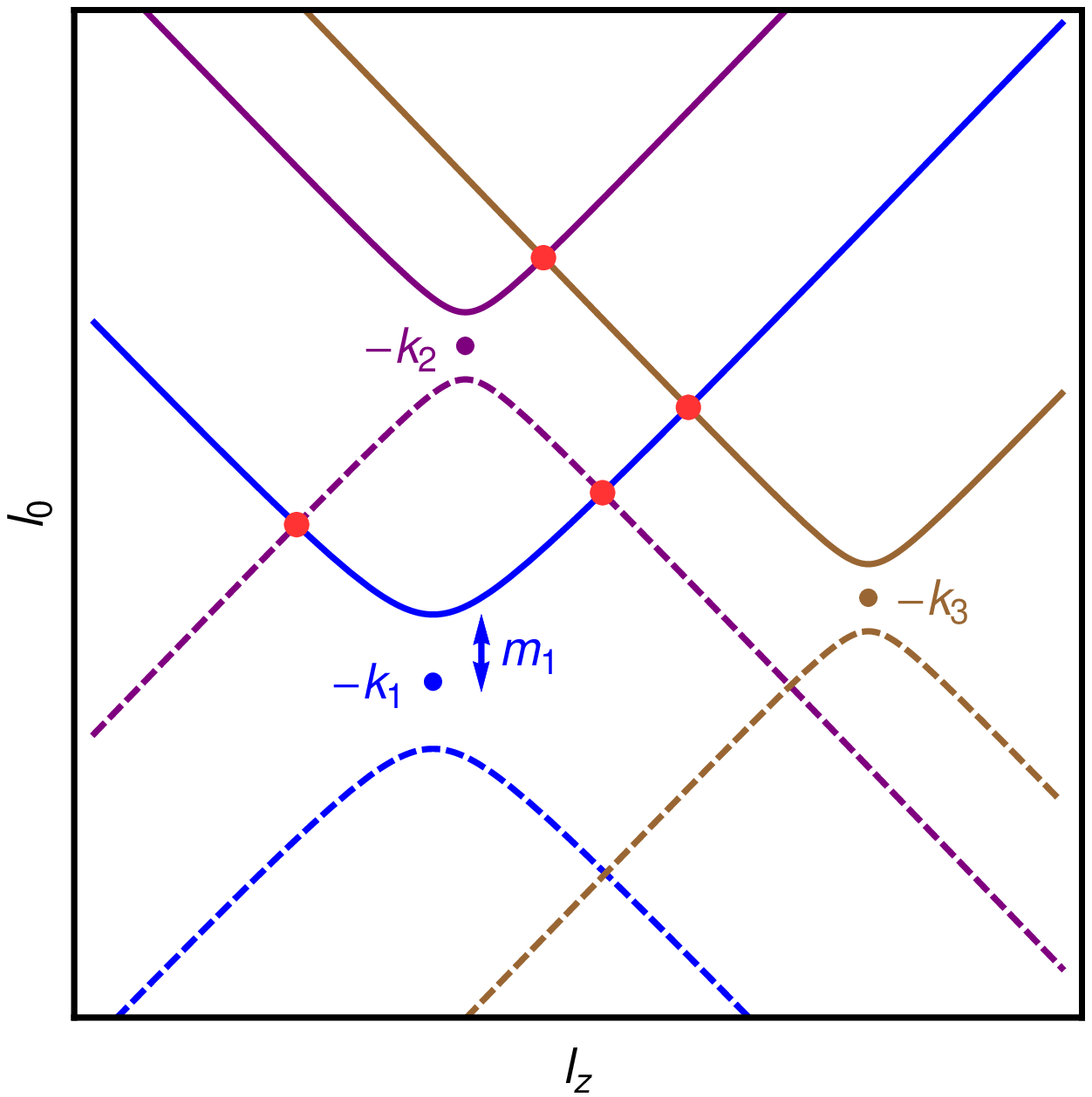}
\includegraphics[width=6cm]{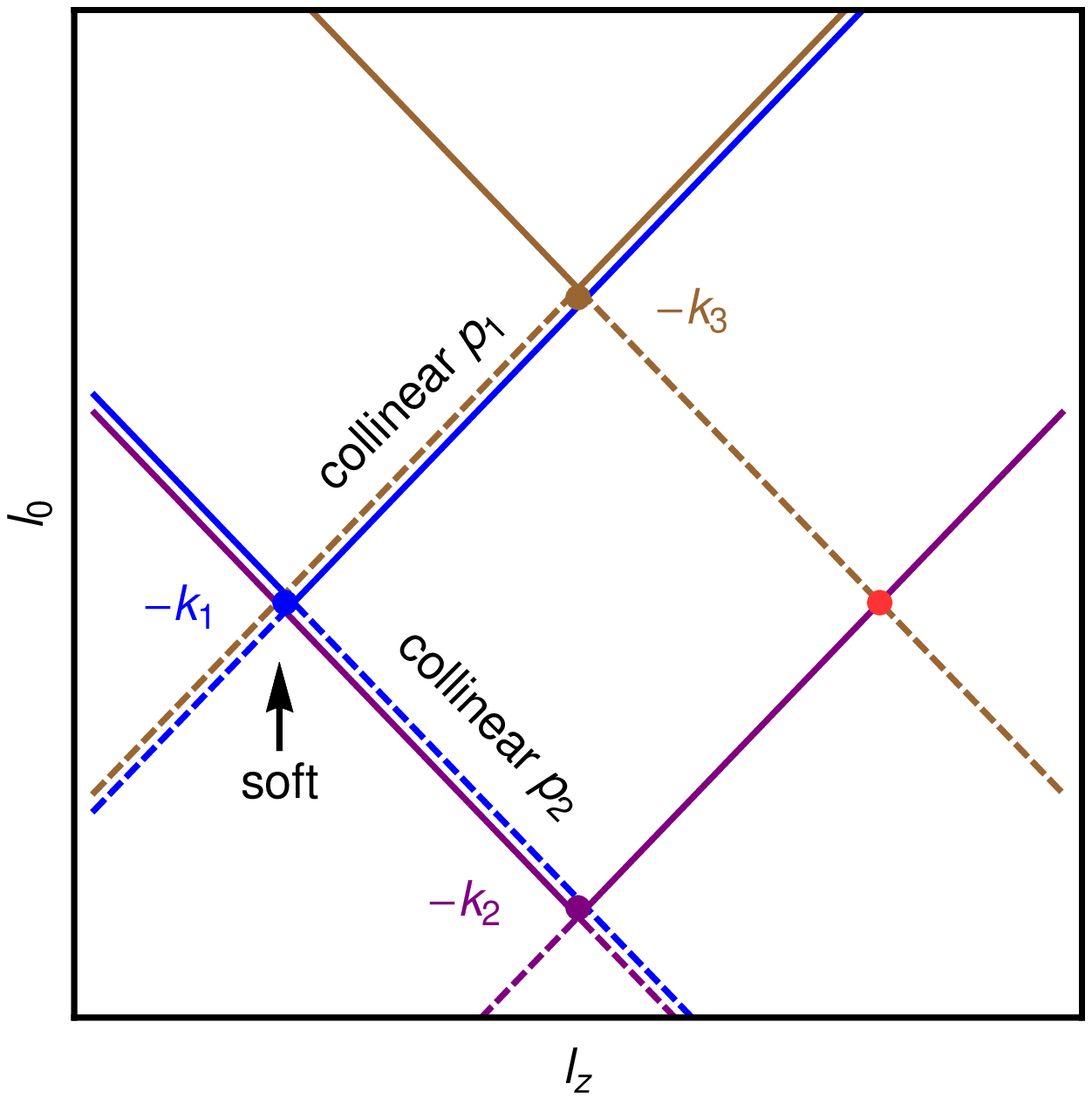}
\caption{Light-cone hyperboloids for three arbitrary propagators 
in Cartesian coordinates in the ($\ell_0$,$\ell_z$) space (left). 
Kinematical configuration with infrared singularities (right). 
\label{fig:cartesean}}
\end{center}
\end{figure}
%%%%%%%%%%%%%%%

Analysing the singular behaviour of the loop integrand 
in the loop momentum space is an attractive approach because it allows a 
rather direct physical interpretation of the singularities 
of the loop quantities~\cite{Sterman:1978bi}. This is particularly 
true for the case of the loop-tree duality method.
The loop integrand becomes singular in regions of the 
loop momentum space in which subsets of internal lines go on-shell.
In Cartesian coordinates, the Feynman propagator in~\Eq{eq:feynman}  
becomes singular at hyperboloids with origin in $-k_{i}$, 
where the minimal distance between each hyperboloid and 
its origin is determined by the internal mass $m_i$.
This is illustrated in Fig.~\ref{fig:cartesean}, where for simplicity
we work in $d=2$ space-time dimensions. Figure~\ref{fig:cartesean}~(left)
shows a typical kinematical situation where two 
momenta, $k_1$ and $k_2$, are separated by a time-like distance, 
$k_{21}^2 > 0$, and a third momentum, $k_3$, is space-like separated  
with respect to the other two, 
$k_{31}^2 <0$ and $k_{32}^2 <0$. The forward light-cone hyperboloids
($q_{i,0}>0$) are represented in Fig.~\ref{fig:cartesean} by solid lines, 
and the backward light-cone hyperboloids ($q_{i,0}<0$) by dashed lines. 
The loop-tree duality method is equivalent to performing the 
loop integration along the forward light-cone hyperboloids.
In the following, we take $\eta_\mu = (1,\mathbf{0})$, and thus 
$- i0 \, \eta \, k_{ji} = -i0 \, k_{ji,0}$. 

Two or more Feynman propagators become simultaneously singular 
where their respective light-cone hyperboloids intersect. 
In most cases, these singularities, due to normal or anomalous 
thresholds~\cite{Mandelstam:1960zz,Rechenberg:1972rq}
of intermediate states, are integrable 
by contour deformation~\cite{Gong:2008ww}.
However, if two massless propagators are separated by a 
light-like distance, $k_{ji}^2=0$, then the overlap is tangential, 
as illustrated in Fig.~\ref{fig:cartesean}~(right),
and leads to non-integrable collinear singularities. 
In addition, massless propagators can generate soft singularities 
at $q_{i}=0$. 
In the dual representation of the integrand at least one propagator 
is already set on-shell, and we should analyse the singularities 
of the dual propagators.  
A crucial point of our discussion is the observation that 
dual propagators can be rewritten as 
\beq
\td{q_i} \, G_D(q_i;q_j) = i\, 2 \pi \, 
\frac{\delta(q_{i,0}-q_{i,0}^{(+)})}{2 q_{i,0}^{(+)}} \, 
\frac{1}{(q_{i,0}^{(+)} + k_{ji,0})^2-(q_{j,0}^{(+)})^2}~,
\label{eq:newdual}
\eeq
where
\beq
q_{i,0}^{(+)} = \sqrt{\mathbf{q}_i^2 + m_i^2-i0}
\label{qi0distance}
\eeq
is the loop energy measured along the light-cone hyperboloid 
with origin at $-k_i$. By definition we have ${\rm Re}(q_{i,0}^{(+)}) \ge 0$. 
The factor $1/q_{i,0}^{(+)}$ can become singular for $m_i=0$, but the integral 
$\int_\ell \delta(q_{i,0}-q_{i,0}^{(+)})/q_{i,0}^{(+)}$ is still convergent by two 
powers in the infrared. For soft singularities to be generated
two dual propagators, each one contributing with one power 
in the infrared, are required. 
From \Eq{eq:newdual} it is obvious that dual propagators become 
singular, $G_D^{-1}(q_i;q_j)=0$, if one of the following conditions is fulfilled:
\bea
&& q_{i,0}^{(+)}+q_{j,0}^{(+)}+k_{ji,0}=0~, \label{ellipsoid} \\
&& q_{i,0}^{(+)}-q_{j,0}^{(+)}+k_{ji,0}=0~. \label{hyperboloid}
\eea
The first condition, \Eq{ellipsoid}, is satisfied if the 
forward light-cone hyperboloid of $-k_i$ intersects 
with the backward light-cone hyperboloid of $-k_j$:
\beq
k_{ji}^2-(m_j+m_i)^2 \ge 0~, \qquad k_{ji,0}<0~, \qquad 
\rm{forward~with~backward~light-cones}~.
\label{eq:generalizedtimelike}
\eeq
The second condition, \Eq{hyperboloid}, is true when 
the two forward light-cone hyperboloids intersect each other: 
\beq
k_{ji}^2-(m_j-m_i)^2 \le 0~, \qquad \rm{two~forward~light-cones}~.
\label{eq:generalizedspacelike}
\eeq

One of the main properties and advantages of the loop-tree 
duality method is that a partial cancellation of 
singularities occurs among different dual integrands~\cite{Buchta:2014dfa}.
For a qualitative discussion, 
let's go back to Fig.~\ref{fig:cartesean}~(left)
where two of the Feynman propagators are separated by 
a space-like distance, $k_{31}^2 < 0$ (or more generally 
fulfilling~\Eq{eq:generalizedspacelike}). 
In the corresponding dual representation one of these
propagators is set on-shell while the other becomes dual, 
and the integration occurs along the respective 
forward light-cone hyperboloids.
There, the two forward light-cone hyperboloids of $-k_1$ and $-k_3$ intersect 
at a single point. The integration over $\ell_z$ along the forward 
light-cone hyperboloid of $-k_1$ occurs outside the forward light-cone 
hyperboloid of $-k_3$ below the singular intersection point, 
and inside that light-cone above this point. 
The opposite occurs if we set $q_3$ on-shell; 
integration over $\ell_z$ happens from inside to outside the 
forward light-cone hyperboloid of $-k_1$.
This leads to a change of sign and to the cancellation of the 
common singularity between the two dual contributions.
Similarly, three and four space-like separated propagators 
do not lead to any common singularity.
For a detailed analytic demonstration see Ref.~\cite{Buchta:2014dfa}.
If instead, the separation is time-like (in the sense 
of~\Eq{eq:generalizedtimelike}), as is the case of $k_2$ with 
respect to $k_1$ in Fig.~\ref{fig:cartesean}~(left), 
the common singularities are met only by one of the two forward 
light-cone hyperboloids, and then only one of the two dual integrands 
becomes singular.  

A similar qualitative analysis is extensible to collinear singularities, 
occurring when two massless propagators are separated by a light-like 
distance, e.g. $k_{31}^2=0$ in Fig.~\ref{fig:cartesean}~(right).
In that case, the corresponding light-cone hyperboloids overlap 
tangentially along an infinite interval. 
The collinear singularity for $\ell_0 > -k_{3,0}$, however, appears 
at the intersection of the two forward light-cone hyperboloids, 
with the forward light-cone hyperboloid of $-k_{3}$ located
inside the forward light-cone hyperboloid of $-k_{1}$, equivalently 
with the forward light-cone hyperboloid of $-k_{1}$ located
outside the forward light-cone hyperboloid of $-k_{3}$, and then 
the singularities cancel each other. 
For $-k_{1,0} < \ell_0 < -k_{3,0}$, instead, it is the forward 
light-cone hyperboloid of $-k_1$ that intersects tangentially 
with the backward light-cone hyperboloid of $-k_3$ according 
to~\Eq{ellipsoid}. 
The collinear divergences survive in this energy strip, 
which indeed also limits the range of the loop three-momentum. 
The soft singularity of the integrand
at $q_{i,0}^{(+)}=0$ leads to soft divergences only if two other 
propagators, each one contributing with one power in the infrared, 
are light-like separated from $-k_i$. In Fig.~\ref{fig:cartesean}~(right)
this condition is fulfilled at $q_{1,0}^{(+)}=0$ only.

%%%%%%%%%%%%%%%
\begin{figure}[ht]
\begin{center}
\includegraphics[width=13cm]{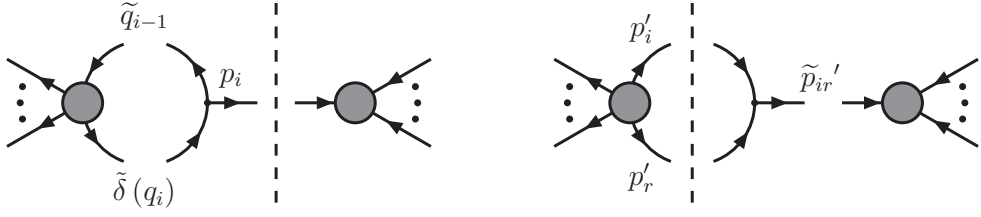}
\caption{\label{fig:collinear} 
{\em Factorization of the dual one-loop and tree-level squared amplitudes 
in the collinear limit. The dashed line represents the momentum conservation
cut.}}
\end{center}
\end{figure}
%%%%%%%%%%%%%%%

\section{Cancellation of infrared singularities with real corrections}
\label{sec:real}

In the previous section we have seen that both threshold and 
infrared singularities are constrained in the dual representation 
of the loop integrand to a finite region of the loop three-momentum.
Singularities outside this region, 
occurring in the intersection of forward light-cone hyperboloids, 
cancel in the sum of all the dual contributions. 
The size of this region is of the order of the external momenta, 
and can be mapped to the finite-size phase-space of the real corrections. 
To discuss the cancellation of infrared singularities 
with the real corrections we make use of collinear factorization 
and the splitting matrices, which encode the collinear singular 
behaviour of scattering amplitudes, as introduced in Ref.~\cite{Catani:2003vu}. 
We consider the one-loop scattering amplitude ${\cal M}^{(1)}_N$
with the internal momenta $q_i$ on-shell and 
the limit where $\pb_i$ and $\qb_i$ become collinear 
\bea
\label{eq:piqicoll}
\ket{{\cal M}^{(0)}_{N} (p_1, \ldots, p_{N})}  
&\to& \ket{{\cal M}^{(0)}_{N+2} (\ldots, p_i, -q_i, q_i, p_{i+1}, \ldots)} 
\\ &=&
\sp^{(0)}(p_i, -q_i; -\widetilde{q}_{i-1}) \, 
\ket{\overline{\cal M}^{(0)}_{N+1} 
(\ldots, p_{i-1}, -\widetilde{q}_{i-1}, q_i, p_{i+1}, \ldots)} 
+ {\cal O}(q_{i-1}^2)~, \nn
\eea
where the reduced matrix element $\overline{\cal M}^{(0)}_{N+1}$
is obtained by replacing the two collinear partons of 
${\cal M}^{(0)}_{N+2}$ by a single parent parton with light-like 
momentum $\widetilde{q}_{i-1}^\mu = q_{i-1}^\mu - 
\frac{q_{i-1}^2 \, n^\mu}{2  \, n q_{i-1}}$, with $n^2=0$. 
Its interference with the corresponding $N$-parton 
tree-level scattering amplitude ${\cal M}^{(0)}_N$, 
is integrated with the appropriate phase-space factor 
\beq
\int d\Phi_{N}(p_1; p_2,\ldots, p_N) = 
\left(\prod_{i=2}^N \int_{p_i} \td{p_i} \right) \, 
\delta^{(d)} (\sum_{i=1}^{N} p_i) \, \theta(p_{i,0}-q_{i,0}^{(+)})~,
\label{phasespace}
\eeq
where we assume that only the external momentum $p_1$ is incoming ($p_{1,0}<0$).
Notice that the loop energy in \Eq{phasespace} is restricted by the 
energy of the external particle $p_i$ because we have selected the infrared 
divergent region. This restriction allows for the mapping with 
real corrections, as illustrated in \Eq{fig:collinear}.  

Similarly, we consider the $N+1$ tree-level scattering amplitude where the 
parton $i$ radiates an extra parton $r$. Besides the initial state momentum 
$p_1$, we denote the external momenta of the real corrections as primed 
momenta because they are subject to the momentum conservation delta function.
In the limit where $\pb_i'$ and $\pb_{r}'$ become collinear, 
${\cal M}^{(0)}_{N+1}$ factorizes as 
\beq
\bra{{\cal M}^{(0)}_{N+1} (p_1, \ldots, p_{N+1}')} =
\bra{\overline{\cal M}^{(0)}_N 
(\ldots, p_{i-1}', \widetilde{p}_{ir}', p_{i+1}', \ldots)} \, 
\sp^{(0) \dagger}(p_i', p_{r}'; \widetilde{p}_{ir}') 
+ {\cal O}(s_{ir}')~, 
\label{eq:piprimecoll}
\eeq
where $p_{ir}' = p_{i}'+p_{r}'$, $s_{ir}' = p_{ir}'^2$, and
$\widetilde{p}_{ir}'^{\mu} = p_{ir}'^{\mu} 
- \frac{s_{ir}' \, n^\mu}{2\, n  p_{ir}'}$.
As Fig.~\ref{fig:collinear} suggests the mapping between 
the four-momenta of the virtual and real matrix 
elements should be such that 
$p_i = \widetilde{p}_{ir}'$,
$p_{j} = p_{j}' (j \ne i)$,
$-\widetilde{q}_{i-1} = p_i'$ and 
$q_i = p_{r}'$,
in the collinear limit. For more details see Ref.~\cite{Buchta:2014dfa}.

\section{Conclusions and outlook}
\label{sec:conclusions}

The loop-tree duality method presents quite attractive features 
for the calculation of multipartonic cross-sections at higher orders.
Integrand singularities occurring in the intersection 
of forward light-cones, or equivalently from space-like separated propagators, 
cancel among dual integrals. 
The remaining singularities, excluding UV divergences,
are found in the intersection of forward with backward light-cones and are 
produced by dual propagators that are time-like separated 
(or causally connected) and less energetic than the internal 
propagator that is set on-shell. Therefore, these singularities 
are restricted to a finite region of the loop three-momentum space, 
which is of the size of the external momenta. As a result, 
a local mapping at the integrand level is possible between one-loop 
and tree-level matrix elements to cancel soft and collinear divergences. 
One can anticipate that a similar analysis at higher orders of the 
loop-tree duality relation is expected to provide equally interesting results.

{\it Acknowledgements:} We thank S. Catani for 
a fruitful long-term collaboration. 
This work has been supported by the EU
under the Grant Agreement PITN-GA-2010-264564 (LHCPhenoNet),
the Spanish Government and EU ERDF funds 
(FPA2011-23778 and CSD2007-00042 
CPAN) and GV (PROMETEUII/2013/007). 
SB acknowledges support from JAEPre (CSIC).
GC from Marie Curie Actions (PIEF-GA-2011-298582).
IB acknowledges support from the German Science Foundation
(DFG) within the Collaborative Research Centre 676
``Particles, Strings and the Early Universe''.


\begin{thebibliography}{99}

%%%% loop--tree duality %%%%

\bibitem{Catani:2008xa}
  S.~Catani, T.~Gleisberg, F.~Krauss, G.~Rodrigo and J.~C.~Winter,
  %``From loops to trees by-passing Feynman's theorem,''
  JHEP {\bf 0809} (2008) 065.
  %[arXiv:0804.3170 [hep-ph]].
  %%CITATION = ARXIV:0804.3170;%%

\bibitem{Bierenbaum:2010cy}
  I.~Bierenbaum, S.~Catani, P.~Draggiotis and G.~Rodrigo,
  %``A Tree-Loop Duality Relation at Two Loops and Beyond,''
  JHEP {\bf 1010} (2010) 073.
  %[arXiv:1007.0194 [hep-ph]].
  %%CITATION = ARXIV:1007.0194;%%

\bibitem{Bierenbaum:2012th}
  I.~Bierenbaum, S.~Buchta, P.~Draggiotis, I.~Malamos and G.~Rodrigo,
  %``Tree-Loop Duality Relation beyond simple poles,''
  JHEP {\bf 1303} (2013) 025.
  %[arXiv:1211.5048 [hep-ph]].
  %%CITATION = ARXIV:1211.5048;%%
%\bibitem{Bierenbaum:2013nja}
  I.~Bierenbaum %, P.~Draggiotis, S.~Buchta, G.~Chachamis, I.~Malamos and G.~Rodrigo,
  %``News on the Loop-tree Duality,''
  {\it et al.},
  Acta Phys.\ Polon.\ B {\bf 44} (2013) 11,  2207.
  %%CITATION = APPOA,B44,2207;%%

\bibitem{Buchta:2014dfa}
  S.~Buchta, G.~Chachamis, P.~Draggiotis, I.~Malamos and G.~Rodrigo,
  %``On the singular behaviour of scattering amplitudes in quantum field theory,''
  arXiv:1405.7850 [hep-ph].
  %%CITATION = ARXIV:1405.7850;%%
 
\bibitem{Feynman:1963ax}
  R.~P.~Feynman,
  %``Quantum theory of gravitation,''
  Acta Phys.\ Polon.\  {\bf 24} (1963) 697;
  %%CITATION = APPOA,24,697;%%
%\bibitem{F2}
%  R.~P.~Feynman,
%  {\it Closed Loop And Tree Diagrams,}
%  in {\it Magic Without Magic}, ed. J.~R.~Klauder, 
%  (Freeman, San Francisco, 1972), p.~355, 
%  in {\it Selected papers of Richard Feynman}, %ed.
%  L.~M.~Brown (World Scientific, Singapore, 2000) p.~867. 

\bibitem{Sterman:1978bi}
  G.~F.~Sterman,
  %``Mass Divergences in Annihilation Processes. 1. Origin and Nature of Divergences in Cut Vacuum Polarization Diagrams,''
  Phys.\ Rev.\ D {\bf 17} (1978) 2773.
  %%CITATION = PHRVA,D17,2773;%%

\bibitem{Gong:2008ww}
  D.~E.~Soper,
  %``QCD calculations by numerical integration,''
  Phys.\ Rev.\ Lett.\  {\bf 81} (1998) 2638.
  %[hep-ph/9804454].
  %%CITATION = HEP-PH/9804454;%%

\bibitem{Mandelstam:1960zz}
  S.~Mandelstam,
  %``Unitarity Condition Below Physical Thresholds in the Normal and Anomalous Cases,''
  Phys.\ Rev.\ Lett.\  {\bf 4} (1960) 84. 
  %%CITATION = PRLTA,4,84;%%

\bibitem{Rechenberg:1972rq}
  H.~Rechenberg and E.~C.~G.~Sudarshan,
  %``Analyticity in quantum field theory. 1. The triangle graph revisited,''
  Nuovo Cim.\ A {\bf 12} (1972) 541.
  %%CITATION = NUCIA,A12,541;%%

%%% 
\bibitem{Chetyrkin}
  K.~G.~Chetyrkin and F.~V.~Tkachov,
  %``Integration by Parts: The Algorithm to Calculate beta Functions in 4 Loops,''
  Nucl.\ Phys.\ B {\bf 192} (1981) 159.
  %%CITATION = NUPHA,B192,159;%%

\bibitem{Smirnov}
V.A. Smirnov, \textit{Feynman Integral Calculus}, Springer-Verlag (2006)

\bibitem{Catani:2003vu}
  S.~Catani, D.~de Florian and G.~Rodrigo,
  %``The Triple collinear limit of one loop QCD amplitudes,''
  Phys.\ Lett.\ B {\bf 586} (2004) 323.
  %[hep-ph/0312067].
  %%CITATION = HEP-PH/0312067;%%
%\bibitem{Sborlini:2013jba}
  G.~F.~R.~Sborlini, D.~de Florian and G.~Rodrigo,
  %``Double collinear splitting amplitudes at next-to-leading order,''
  JHEP {\bf 1401} (2014) 018.
  %[arXiv:1310.6841 [hep-ph]].
  %%CITATION = ARXIV:1310.6841;%%
 
\end{thebibliography}
\end{document}